\documentclass[journal]{IEEEtran}

\usepackage{graphicx}
\usepackage{amsmath}
\usepackage{amssymb}
\usepackage{textcomp}
\usepackage{url}
\usepackage[T1]{fontenc}

\begin{document}

\title{Pattern Multiplication Underestimates the Sidelobe Level of a Planar
Slotted-Waveguide Array by 1.1 to 2.8~dB Across a 2\,\% Band}

\author{William~Khalili~Jr.%
\thanks{The author is an independent researcher. (e-mail:
william.khalilijr@gmail.com).}}

\markboth{IEEE Transactions on Antennas and Propagation}%
{Khalili: Pattern Multiplication Underestimates Slotted-Waveguide Array Sidelobes}

\maketitle

\begin{abstract}
The standard shortcut in slotted-waveguide array design---solve one radiating
stick, multiply its embedded pattern by the array factor of the intended
aperture distribution, and accept the product as the array pattern---is tested
against a full-wave solution of the same metal. For a $16\times16$-slot planar
array in WR-90 synthesised to a $-30$~dB Taylor illumination, the shortcut
underestimates the transverse-plane sidelobe level by $1.06$~dB at the design
frequency and by $2.22$ and $2.84$~dB at the edges of a 2\,\% band: a spread of
$1.78$~dB, larger than the mid-band error itself. Every sidelobe figure carries
$\pm0.11$~dB of mesh spread from a convergence study on the production
geometry. The error is signed, is worst at the band edges where the design has
least margin, and falls only from $1.58$~dB to $1.06$~dB when the aperture is
doubled from eight sticks to sixteen, which identifies the edge elements as its
source. The array was made analysable by a thirteen-point single-slot full-wave
calibration: over the design's offset range the handbook closed form for
resonant length moves $3.9~\mu$m where the measurement moves $143.9~\mu$m, and
Stevenson's conductance is shown to overestimate by a nearly constant
$5.16$\,\%.
\end{abstract}

\begin{IEEEkeywords}
Antenna arrays, slot antennas, waveguide antennas, mutual coupling, sidelobe
level, electromagnetic modeling, monopulse, antenna measurements.
\end{IEEEkeywords}

\IEEEpeerreviewmaketitle

% =====================================================================
\section{Introduction}
% =====================================================================
\IEEEPARstart{T}{he} longitudinal shunt slot cut in the broad wall of a
rectangular waveguide is the oldest and still the most-used radiating element
in planar microwave arrays. Its design theory is settled in outline. Watson
established the equivalent-circuit description---a slot is a lumped element in
a transmission line, an array is the cascade of those elements
\cite{watson1947}; Stevenson supplied a closed form for the shunt conductance
by a power-balance argument \cite{stevenson1948}; Oliner removed the
zero-thickness and zero-width assumptions variationally \cite{oliner1957}; and
Elliott addressed the fact that a slot in an array is not a slot alone
\cite{elliott1978,elliott1979,elliott1981}. Modern treatments of the whole
subject are available \cite{josefsson2018,silver1949}.

What is not settled is the step every one of those designs must take at the
end, when the single-guide theory has produced a set of offsets and lengths and
the designer wants the pattern of the finished planar aperture. The near
universal practice is \emph{pattern multiplication}: solve one stick full-wave,
take its embedded pattern, multiply by the array factor of the intended
inter-stick distribution, and report the product. The step is attractive for a
reason that has nothing to do with its accuracy---a single stick is cheap to
solve and a sixteen-port solution of the whole aperture is not---and its
justification is usually left implicit, resting on the well-established
observation that the \emph{active} element pattern of a large array is nearly
uniform across its interior \cite{pozar1994}.

The gap this paper addresses is that the size of the error committed by that
step, for a real slotted-waveguide aperture, has not been reported as a
measured quantity with its frequency dependence and its numerical uncertainty
attached. Sidelobe level is precisely the quantity the shortcut is used to
claim, precisely the quantity a slot array is chosen for, and precisely the
kind of quantity---one formed by near-cancellation---that is most exposed to a
missing piece of physics.

This paper supplies that number. The contributions are:

\begin{enumerate}
\item A direct measurement of the pattern-multiplication error on one planar
slotted-waveguide array, at three frequencies spanning 2\,\% of bandwidth, with
the mesh uncertainty of every figure stated. The error is
$+1.06$ to $+2.84$~dB in sidelobe level, always signed the same way, and its
$1.78$~dB spread across the band is larger than its value at mid-band.
\item Evidence that the error is carried by the edge elements: the same
measurement on an eight-stick array of the same sticks gives $+1.58$~dB against
$+1.06$~dB on sixteen, the right sign and order for a fault carried by a fixed
number of elements while the aperture grows.
\item A thirteen-point single-slot full-wave calibration that replaces two
closed forms, together with the demonstration that one of them---the handbook
resonant-length expression---carries essentially no information over the range
in which it is used, and that the other, Stevenson's conductance, is wrong in
value by a nearly constant factor and right in shape.
\end{enumerate}

Section~\ref{sec:article} describes the test article and states why it was
synthesised rather than borrowed. Section~\ref{sec:cal} gives the two-port
extraction and the calibration. Section~\ref{sec:synth} gives the array
synthesis and the built geometry, Section~\ref{sec:fw} the full-wave method and
its convergence, Section~\ref{sec:results} the result, and
Section~\ref{sec:discussion} what a designer should do instead.

% =====================================================================
\section{The Test Article, and Why It Was Synthesised}
\label{sec:article}
% =====================================================================
The object of study is sixteen slotted waveguide sticks in WR-90, sixteen
longitudinal shunt slots in each, standing side by side so that the sticks
touch. It was synthesised for this work rather than taken from an existing
design, and the reason is the claim being made. \textbf{A test piece whose
target aperture distribution is known exactly is a stronger object of study
than a borrowed one whose target is unknown}, because only the former can
support a statement of the form \emph{the full-wave result departs from the
intended distribution by this much}; a borrowed article can be measured but not
audited.

Table~\ref{tab:geom} collects the geometry. The guide is WR-90,
$a = 22.86$~mm and $b = 10.16$~mm internally, with a broad wall $t = 1.27$~mm
thick. The design frequency is $f_0 = 9.375$~GHz, where
$\lambda_0 = 31.9779$~mm and the guide wavelength is $\lambda_g = 44.7429$~mm.
Slots are spaced $\lambda_g/2 = 22.3714$~mm along each stick with offsets
alternating about the centreline, the feed reaches the first slot one guide
wavelength in, and a short circuit stands $\lambda_g/4$ beyond the last, giving
a stick $391.50$~mm long---$8.75$ guide wavelengths. The intended illumination
is a Taylor distribution \cite{taylor1955} with $\bar{n}=5$ at $-30$~dB,
applied in both planes.

Two consequences of the mechanical arrangement should be read before any
pattern is. First, because the sticks touch, the lattice across the aperture is
the guide's \emph{outside} width, $25.4$~mm, which is $0.7943\lambda_0$ at
$f_0$; no electrical decision can change it, and it limits the grating-lobe-free
scan to $13.27^\circ$ at the top of the band. Second, the aperture is
$406.40 \times 357.94$~mm, an area of $0.145468$~m$^2$, whose directivity
ceiling $4\pi A/\lambda_0^2$ is $32.523$~dBi.

\begin{figure*}[!t]
\centering
\includegraphics[width=\textwidth]{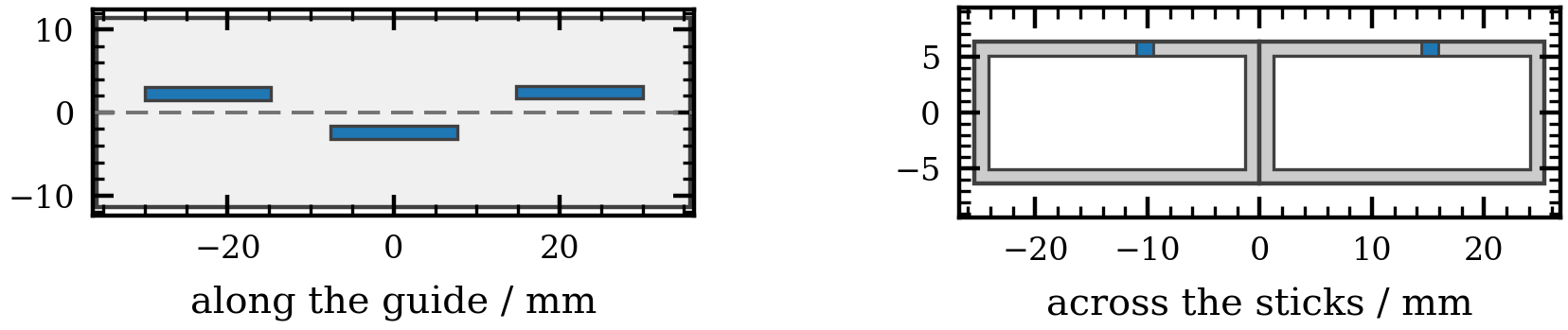}
\caption{The test article, drawn to scale from the synthesis of
Section~\ref{sec:synth} rather than sketched; both axes of both panels are in
millimetres. Left: three slots near the middle of a stick, at a pitch of
$\lambda_g/2 = 22.37$~mm, with offsets alternating about the centreline
(dashed). Right: two adjacent sticks in section. The sticks touch, so the
lattice across the aperture is the guide's outside width, $25.4$~mm, which is
$0.7943\lambda_0$ at the design frequency.}
\label{fig:geometry}
\end{figure*}

Fig.~\ref{fig:geometry} shows the array drawn to scale from the synthesis of
Section~\ref{sec:synth}.

\begin{table}[!t]
\renewcommand{\arraystretch}{1.15}
\caption{The Test Article}
\label{tab:geom}
\centering
\begin{tabular}{lr}
\hline
Quantity & Value \\
\hline
Waveguide & WR-90 \\
Broad wall $a$ (internal) & $22.86$~mm \\
Narrow wall $b$ (internal) & $10.16$~mm \\
Broad-wall thickness $t$ & $1.27$~mm \\
Design frequency $f_0$ & $9.375$~GHz \\
Free-space wavelength $\lambda_0$ & $31.9779$~mm \\
Guide wavelength $\lambda_g$ & $44.7429$~mm \\
Slots per stick & 16 \\
Slot pitch along a stick, $\lambda_g/2$ & $22.3714$~mm \\
Sticks & 16 \\
Stick pitch (guide outside width) & $25.4$~mm ($0.7943\lambda_0$) \\
Stick length & $391.50$~mm ($8.75\lambda_g$) \\
Aperture & $406.40 \times 357.94$~mm \\
Intended illumination & Taylor, $\bar{n}=5$, $-30$~dB \\
Slot offsets, built & $0.6335$ to $2.4499$~mm \\
Slot lengths, built & $15.1044$ to $15.2580$~mm \\
Aperture ceiling $4\pi A/\lambda_0^2$ & $32.523$~dBi \\
\hline
\end{tabular}
\end{table}

% =====================================================================
\section{Single-Slot Calibration}
\label{sec:cal}
% =====================================================================
\subsection{The two-port extraction}
Model one slot as a two-port: a length of guide, a shunt admittance $y$
normalised to the guide's wave admittance, a further length of guide. For the
shunt element alone,
\begin{equation}
S_{11} = -\frac{y}{2+y}, \qquad S_{21} = \frac{2}{2+y},
\label{eq:shunt}
\end{equation}
and the lengths of guide on either side multiply \emph{both} parameters by the
same factor $e^{-2j\beta d}$. Dividing one by the other cancels it:
\begin{equation}
\boxed{\;y = -\,2\,\frac{S_{11}}{S_{21}}\;}
\label{eq:extract}
\end{equation}
This is exact, and it requires knowing neither the length of the guide, nor its
propagation constant, nor the solver's port reference plane. The largest source
of error in this kind of measurement---de-embedding a reference plane that the
simulator places by a convention of its own---is removed by a division.
Equation~(\ref{eq:extract}) was verified numerically before use, by embedding a
known resonant shunt admittance in a deliberately awkward length of dispersive
WR-90 and recovering it: the worst-case recovery error is the arithmetic noise
of double precision, not a tolerance arrived at by adjustment.

The identity assumes what the algebra does not supply: that each port carries
the dominant mode alone. A longitudinal slot is asymmetric across the guide and
scatters TE$_{20}$ as well as TE$_{10}$. In WR-90, TE$_{20}$ cuts off at
$13.114$~GHz, so at $f_0$ it is evanescent with $\alpha = 192.2$~Np/m, which is
$1.669$~dB/mm; $40$~dB of decay therefore takes $24.0$~mm, and the ports in
this campaign stood one guide wavelength, $44.7$~mm, from the slot. A port
placed too close returns a smooth, plausible, repeatable and wrong admittance,
so the stand-off is a number and not a matter of taste. Two checks come free
with the same two parameters: $\mathrm{Re}(y) \ge 0$ at every frequency for a
passive shunt element, and $|S_{11}|^2 + |S_{21}|^2 \le 1$, where the shortfall
is the radiated power.

Resonance was located as the \emph{maximum of the conductance} rather than as
the zero of the susceptance: for the weakest slots of this design the residual
susceptance is of the same order as the conductance itself, so a zero crossing
is poorly conditioned where a peak is not. The peak was interpolated with a
parabola; on a synthetic test this reduced the conductance error from
$0.148$\,\% to $0.0054$\,\%.

\subsection{What the calibration found}
Thirteen offsets were solved, in two stages: a length sweep whose fit predicts
the resonant length, then a run \emph{at} the predicted length, which tests the
prediction. Eight of the thirteen are tabulated in the design record and are
given in Table~\ref{tab:cal}; the fits below were made to all thirteen. The
points span offsets from $0.6079$ to $2.3814$~mm.

\begin{table}[!t]
\renewcommand{\arraystretch}{1.15}
\caption{Single-Slot Calibration, Full-Wave (Eight Tabulated Points of Thirteen)}
\label{tab:cal}
\centering
\begin{tabular}{ccccc}
\hline
Offset & Resonant length & Measured & Stevenson & Ratio \\
$x_1$ / mm & $L$ / mm & $g$ & $g$ & \\
\hline
$0.6079$ & $15.1051$ & $0.00790$ & $0.00860$ & $0.9183$ \\
$0.7649$ & $15.1158$ & $0.01263$ & $0.01360$ & $0.9287$ \\
$1.0483$ & $15.1384$ & $0.02405$ & $0.02546$ & $0.9446$ \\
$1.3977$ & $15.1615$ & $0.04273$ & $0.04502$ & $0.9491$ \\
$1.7443$ & $15.1899$ & $0.06618$ & $0.06963$ & $0.9503$ \\
$2.0404$ & $15.2175$ & $0.08979$ & $0.09461$ & $0.9490$ \\
$2.2605$ & $15.2338$ & $0.10963$ & $0.11543$ & $0.9497$ \\
$2.3814$ & $15.2490$ & $0.12101$ & $0.12765$ & $0.9480$ \\
\hline
\end{tabular}
\end{table}

\emph{The resonant length.} The handbook closed form is
\begin{equation}
L \approx 0.4785\,\lambda_0 - 0.30\,t + 0.012\,\lambda_0 (x_1/a)^2 .
\label{eq:Lclosed}
\end{equation}
Over the eight offsets of Table~\ref{tab:cal} the third term---the only term
that depends on the offset at all---moves the predicted length by
$3.9~\mu$m. The measurement moves it by $143.9~\mu$m, a factor of
thirty-seven. \textbf{Equation~(\ref{eq:Lclosed}) is not a poor prediction of
how resonant length depends on offset; over this range it is not a prediction
at all}---it is a constant with a decorative term attached, and the constant is
itself short, by $184$ to $324~\mu$m across the same range. It was replaced by
a quadratic fitted to all thirteen points,
\begin{equation}
L/\mathrm{mm} = 15.066615 + 0.053136\,\xi + 0.010193\,\xi^2 ,
\label{eq:Lfit}
\end{equation}
with $\xi = x_1/\mathrm{mm}$ and a fit residual of $5.5~\mu$m.

\begin{figure*}[!t]
\centering
\includegraphics[width=0.94\textwidth]{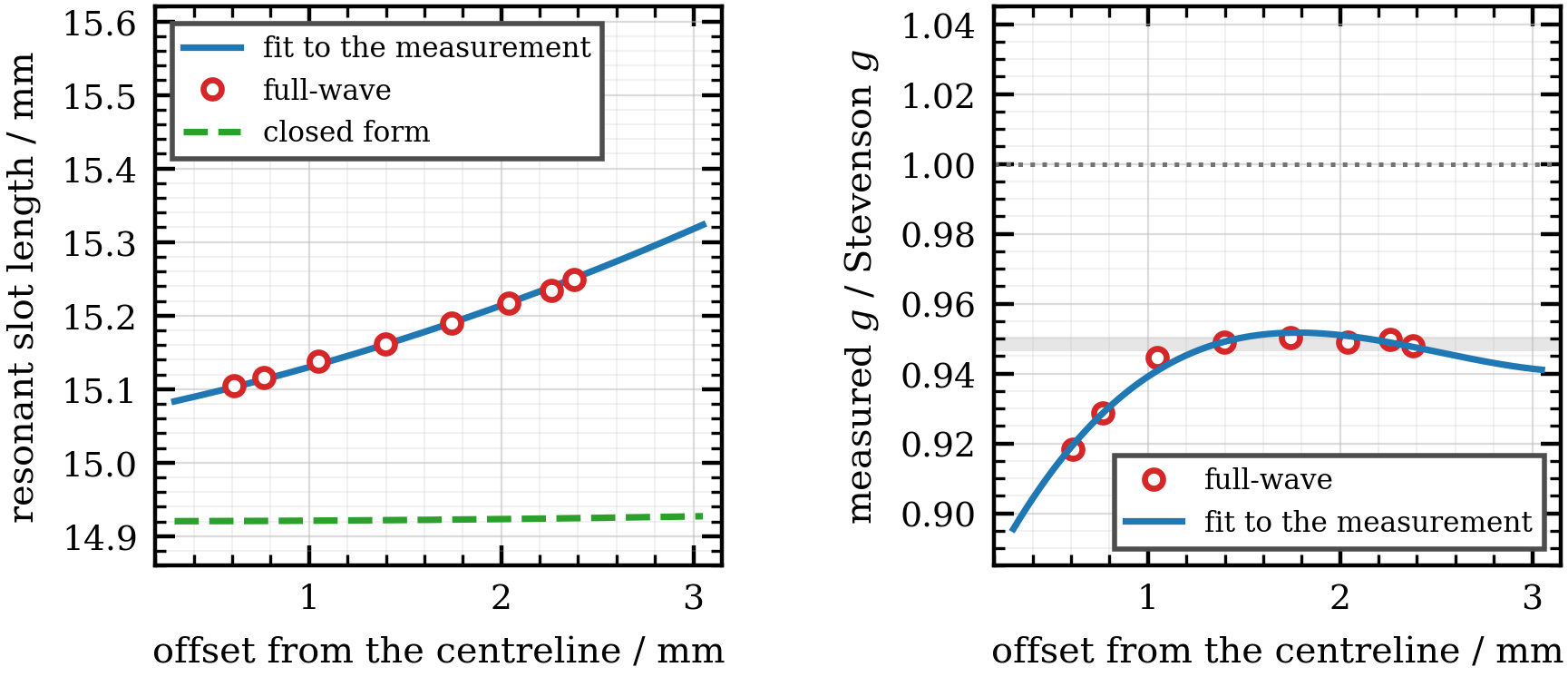}
\caption{The single-slot calibration. Left: the resonant length---full-wave
points, the quadratic of (\ref{eq:Lfit}) fitted to all thirteen, and the closed
form of (\ref{eq:Lclosed}), which is flat on this scale. Right: measured
conductance divided by Stevenson's (\ref{eq:stevenson}), with the shaded band
one standard deviation about the mean of the six points beyond $1$~mm.
Stevenson overestimates by $5.16$\,\% with a scatter of $0.21$\,\% over that
range, so a single constant repairs the law for every slot but the weakest.}
\label{fig:calibration}
\end{figure*}

\emph{The conductance.} Stevenson's law for a longitudinal broad-wall slot,
\begin{equation}
g = 2.09\,\frac{a}{b}\,\frac{\lambda_g}{\lambda_0}
\cos^2\!\left(\frac{\pi\lambda_0}{2\lambda_g}\right)
\sin^2\!\left(\frac{\pi x_1}{a}\right) ,
\label{eq:stevenson}
\end{equation}
reduces at $f_0$ in WR-90 to $K\sin^2(\pi x_1/a)$ with $K = 1.23529$. The
measured-to-Stevenson ratio in the last column of Table~\ref{tab:cal} is
plotted in Fig.~\ref{fig:calibration}. \textbf{Stevenson overestimates, and
over the range this design uses he overestimates by very nearly a single
constant}: above an offset of $1$~mm the ratio is $0.9484 \pm 0.0020$ over six
points---a spread of $0.21$\,\% against a systematic shortfall of
$5.16$\,\%. The mechanism is available: the wall is $0.0397\lambda_0$ thick,
and a slot of finite depth behaves in part like a short section of guide below
cutoff, which attenuates the coupling. A zero-thickness theory cannot know
about it, and Oliner's variational treatment \cite{oliner1957} is why the sign
of the correction was expected before it was measured.

The correction is applied as a \emph{ratio} rather than by fitting the
conductance directly, and the choice matters. A curve fitted to the conductance
itself must span a range of $14.84$ to one and has no data near the origin, so
it goes wrong exactly where the design is most sensitive; a curve fitted to the
ratio spans a few per cent and inherits the correct $g\to0$ limit as
$x_1 \to 0$ from the law it corrects. The fit used is the cubic
\begin{equation}
R(\xi) = 0.863771 + 0.121036\,\xi - 0.052684\,\xi^2 + 0.006987\,\xi^3 ,
\label{eq:ratiofit}
\end{equation}
and the calibrated law is $g(x_1) = R\,g_{\mathrm{Stevenson}}$. It is valid
over the offsets that were measured and nowhere else; a design needing offsets
outside $0.61$--$2.38$~mm needs more slots solved at those offsets, not a
higher order of polynomial.

\subsection{One slot, worked}
The single-slot design is the inversion of the calibrated law. Take the
conductance a uniformly illuminated sixteen-slot stick would need,
$g = 1/16 = 0.0625$. Stevenson alone gives $\sin^2(\pi x_1/a) = g/K =
0.050596$ and hence $x_1 = 1.6509$~mm. The measured ratio there is $R =
0.951438$; inverting once more with $g = KR\sin^2$ gives $x_1 = 1.6932$~mm, and
bisection on the calibrated law to machine precision gives $1.6931$~mm. The
single correction is therefore within $0.1~\mu$m of the converged answer, which
matters because it means the procedure can be done by hand.

The resonant length at that offset, from (\ref{eq:Lfit}), is $15.1858$~mm
$= 0.474885\lambda_0$. The closed form (\ref{eq:Lclosed}) would have given
$14.9225$~mm---short by $263~\mu$m. To first order the fractional shift in a
slot's resonant frequency is the fractional change in its length, so
$263~\mu$m on $15.19$~mm is $1.734$\,\%, which at $9.375$~GHz is $163$~MHz.
\textbf{A stick built to the closed form is tuned to the wrong frequency by
more than its own bandwidth}, which is $146$~MHz to $-15$~dB by full-wave
solution.

% =====================================================================
\section{Array Synthesis and the Built Geometry}
\label{sec:synth}
% =====================================================================
Sixteen slots in one guide, short-circuited $\lambda_g/4$ beyond the last:
the short is an open circuit at the last slot, each slot sits at a voltage
maximum of the standing wave, and slots $\lambda_g/2$ apart sit at successive
maxima, which are in antiphase. Alternating the offsets inverts the sign of the
interrupted wall current and cancels that antiphase exactly, so all sixteen
radiate in phase and the beam is broadside.

Because every slot stands at a voltage maximum of the same standing wave, the
power radiated by slot $n$ is $g_n|V|^2/2$, so the aperture amplitude goes as
the \emph{square root} of the conductance, $a_n \propto \sqrt{g_n}$. Because
the slots are in shunt across one line, the input admittance is their sum, and
requiring a match at $f_0$ fixes the normalisation completely:
\begin{equation}
\sum_{n=1}^{N} g_n = 1, \qquad
g_n = \frac{a_n^2}{\sum_m a_m^2}.
\label{eq:synth}
\end{equation}
These two lines are the whole of the synthesis. Choose the illumination; square
it; normalise it to sum to one; invert the calibrated conductance law for each
offset by bisection; read the resonant length for that offset from
(\ref{eq:Lfit}). Bisection is used rather than a fitted inverse because the law
is monotone over the range a design uses, so bisection is exact, and because it
cannot quietly extrapolate---which is the failure mode that put an eight per
cent error into this campaign's first conductance table.

Table~\ref{tab:stick} is the resulting stick, recomputed here from
(\ref{eq:synth}), (\ref{eq:Lfit}) and (\ref{eq:ratiofit}) rather than
transcribed. The conductances sum to $1.000000000$ by construction, and the
offsets reproduce the built article's, $0.6335$ to $2.4499$~mm, to the
micrometre. The conductance taper spans $14.84$ to one across an offset range
of only $1.82$~mm, which is why the conductance law has to be right rather than
nearly right; that compression is the square-root relation above, and it is why
a slot array's end slots sit so nearly on the centreline.

\begin{figure*}[!t]
\centering
\includegraphics[width=0.94\textwidth]{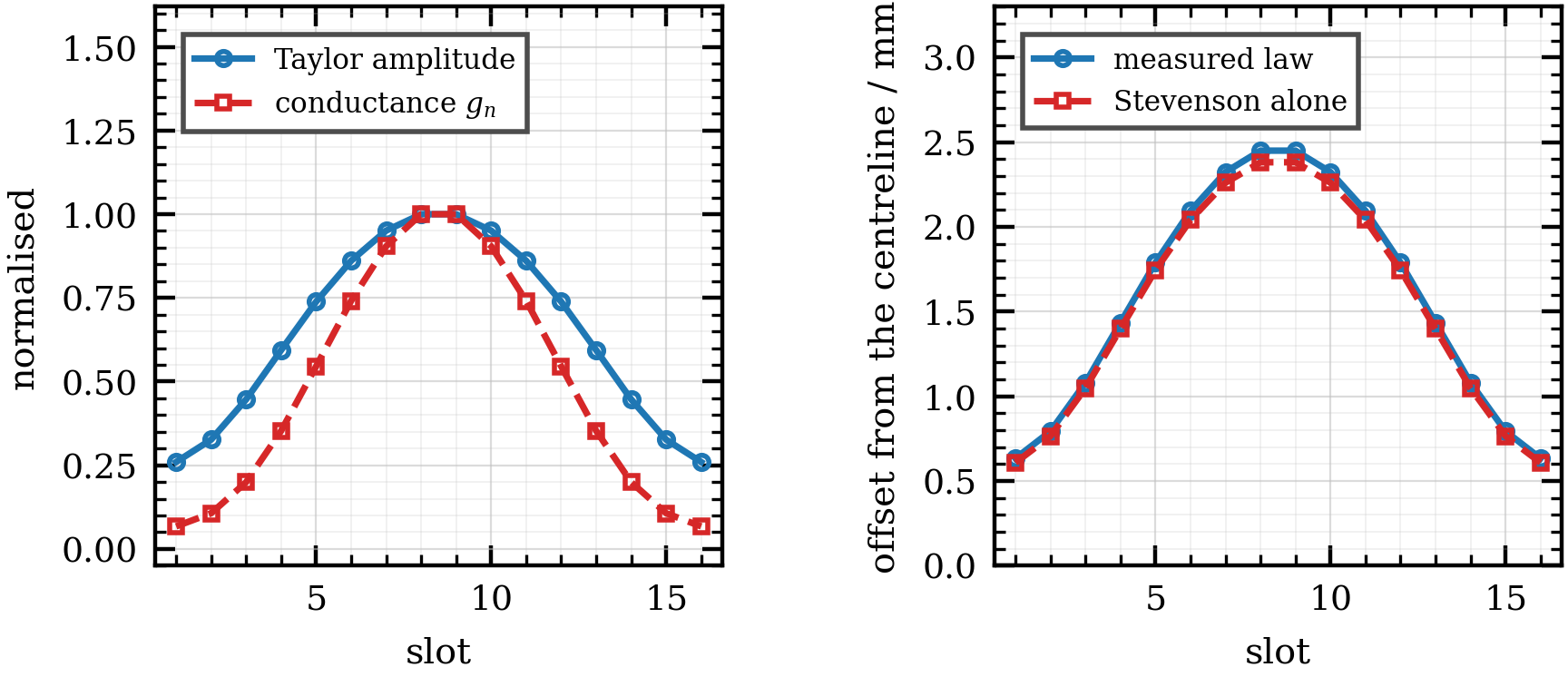}
\caption{Synthesis of the sixteen-slot stick. Left: the Taylor illumination at
$\bar{n}=5$ and $-30$~dB, and the conductances that follow from squaring and
normalising it by (\ref{eq:synth}). The conductance taper is much deeper than
the amplitude taper, which is the square-root relation between them. Right: the
offsets obtained by inverting the calibrated law, against those Stevenson alone
would have given; the two differ by up to $69~\mu$m.}
\label{fig:synthesis}
\end{figure*}

Fig.~\ref{fig:synthesis} shows the synthesis and, in the right panel, the
offsets the calibrated law gives against those Stevenson alone would have
given. The two differ by up to $69~\mu$m.

\begin{table}[!t]
\renewcommand{\arraystretch}{1.15}
\caption{The Synthesised Stick (Slots Are Symmetric About the Centre)}
\label{tab:stick}
\centering
\footnotesize
\setlength{\tabcolsep}{4pt}
\begin{tabular}{ccccr}
\hline
Slot & $g_n$ & $x_1$ / mm & $L$ / mm & Amp. / dB \\
\hline
1, 16 & $0.008602$ & $0.6335$ & $15.1044$ & $-11.71$ \\
2, 15 & $0.013600$ & $0.7932$ & $15.1152$ & $-9.72$ \\
3, 14 & $0.025461$ & $1.0804$ & $15.1359$ & $-7.00$ \\
4, 13 & $0.045018$ & $1.4348$ & $15.1638$ & $-4.53$ \\
5, 12 & $0.069635$ & $1.7889$ & $15.1943$ & $-2.63$ \\
6, 11 & $0.094606$ & $2.0945$ & $15.2226$ & $-1.30$ \\
7, 10 & $0.115427$ & $2.3235$ & $15.2451$ & $-0.44$ \\
8, 9 & $0.127651$ & $2.4499$ & $15.2580$ & $0.00$ \\
\hline
\end{tabular}
\end{table}

\subsection{What the calibration was worth}
\begin{figure*}[!t]
\centering
\includegraphics[width=0.94\textwidth]{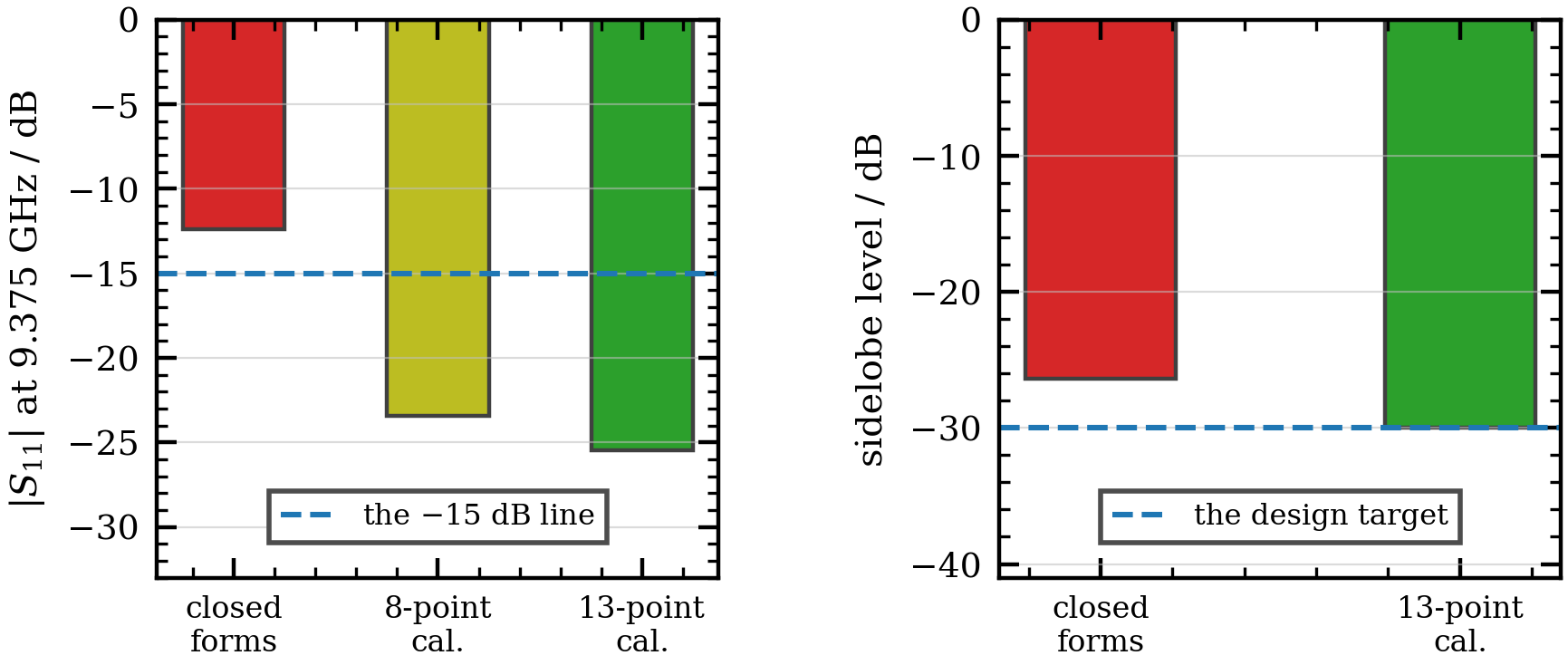}
\caption{The same synthesis built two ways and solved full-wave. Left: the
input match at the design frequency, with the intermediate eight-point
calibration included to show that the sequence converges rather than merely
differing. Right: the sidelobe level in the plane containing the stick axis,
against the design target computed from the array factor of the intended
illumination. The uncalibrated article misses the target by $3.61$~dB; the
calibrated one is within $0.06$~dB.}
\label{fig:bought}
\end{figure*}

The same synthesis was built twice and solved twice: once from the closed forms
(\ref{eq:Lclosed}) and (\ref{eq:stevenson}), and once from the calibrated laws.
A single simulation of the calibrated article would have proved nothing; it is
the pair that measures the calibration, and the uncalibrated article is a
control rather than a straw man---it is what a competent designer following the
handbooks would have cut. Table~\ref{tab:bought} and Fig.~\ref{fig:bought} give
the comparison.

\begin{table}[!t]
\renewcommand{\arraystretch}{1.15}
\caption{One Stick, Built Two Ways, Solved Full-Wave}
\label{tab:bought}
\centering
\begin{tabular}{lccc}
\hline
Quantity & Closed & 13-point & Computed \\
 & forms & calibration & target \\
\hline
$|S_{11}|$ at $9.375$~GHz & $-12.41$~dB & $-25.49$~dB & a match \\
Best $|S_{11}|$ in band & $-13.73$~dB & $-28.38$~dB & --- \\
\quad and where & $9.332$~GHz & $9.358$~GHz & $9.375$~GHz \\
Sidelobe level$^{\dagger}$ & $-26.40$~dB & $-29.95$~dB & $-30.01$~dB \\
Peak directivity & $18.13$~dBi & $18.08$~dBi & --- \\
\hline
\multicolumn{4}{l}{\footnotesize $^{\dagger}$ Carries $\pm0.11$~dB of mesh
spread.}
\end{tabular}
\end{table}

The intermediate eight-point calibration is included in
Fig.~\ref{fig:bought} to show that the sequence converges rather than merely
differing. In sidelobe level the uncalibrated article misses the computed
target by $3.61$~dB and the calibrated one is within $0.06$~dB of it. Note that
the peak directivity barely moves between the two, at $18.13$ against
$18.08$~dBi: \textbf{the calibration is invisible in the quantity most often
quoted and decisive in the two that matter}, the match and the sidelobes.

% =====================================================================
\section{Full-Wave Method and Mesh Convergence}
\label{sec:fw}
% =====================================================================
All full-wave results are from a hexahedral transient solver at 20 lines per
wavelength, $14.76$~M cells for the sixteen-stick array, with every stick fed
at its own port. The array pattern is assembled from the sixteen embedded
patterns and the sixteen-port scattering matrix,
\begin{equation}
E(\theta,\phi) = \frac{\sum_n w_n E_n(\theta,\phi)}
{\sqrt{\,\|w\|^2 - \|\mathbf{S}w\|^2\,}} ,
\label{eq:recon}
\end{equation}
where $E_n$ is the embedded pattern of stick $n$ driven alone and $w$ the
vector of excitation weights. Linearity guarantees the numerator. It does not
guarantee the denominator---the rescaling that undoes the solver's per-port
normalisation, the normalisation of the weights to unit input power, and the
accounting of power reflected back out of the ports. Getting any of those wrong
leaves the pattern \emph{shape} intact while every absolute figure is wrong by
a common factor, which is exactly the kind of error a plot cannot show. The
array was therefore also solved once with all sixteen ports driven together:
the reconstruction and the simultaneous excitation agree to $-0.006$~dB at
$9.200$~GHz, $+0.011$~dB at $9.375$~GHz and $+0.115$~dB at $9.600$~GHz. This
verifies the power bookkeeping, which is the one part of the reconstruction
that linearity does not cover; it does not verify linearity, which was never in
doubt.

An independent check on the absolute level is available from the gain budget.
The aperture ceiling is $32.523$~dBi; the Taylor taper in both planes costs
$1.358$~dB, ohmic loss in the guide $0.024$~dB and the input mismatch
$0.012$~dB, leaving $31.128$~dBi. The full-wave peak at $f_0$ is $31.149$~dBi.

\subsection{Mesh convergence, and what may be quoted}
\begin{figure*}[!t]
\centering
\includegraphics[width=0.94\textwidth]{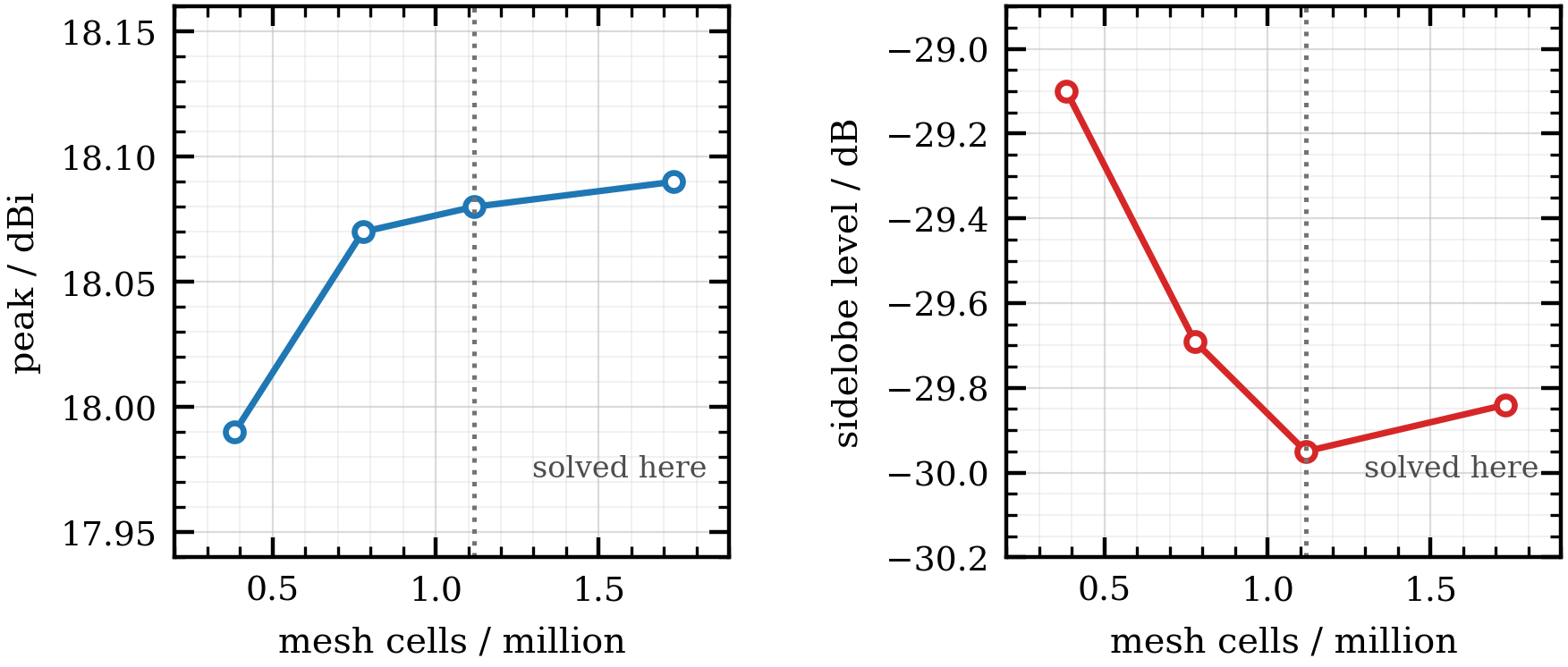}
\caption{Mesh convergence on the production stick, plotted against cell count
rather than the solver's own density setting, which is not comparable between
solvers. Left: the peak directivity, which has converged---the last step moves
it by $0.010$~dB. Right: the sidelobe level, which has not---the last step
moves it by $0.110$~dB and the movement is not monotone, so the figure is a
spread and not a residual trend. The dotted line marks the production density.}
\label{fig:mesh}
\end{figure*}

The production geometry was solved at four mesh densities
(Table~\ref{tab:mesh}, Fig.~\ref{fig:mesh}). The peak directivity is converged:
the last step moves it by $0.010$~dB. \textbf{The sidelobe level is not}: the
last step moves it by $0.110$~dB, and the movement is not monotone, so the
figure is a spread and not a residual trend. Every sidelobe figure in this
paper therefore carries $\pm0.11$~dB.

Two remarks make that number honest. First, a quantity obtained by integrating
over the aperture converges quickly and a quantity obtained by cancellation
does not: the peak is an integral of the aperture field, so a discretisation
error that is random from cell to cell averages down before it reaches the
answer, whereas a sidelobe is the far field at an angle where the contributions
nearly cancel, so the same per-cell error arrives undiminished on a residue
three decades smaller. Second, a mesh is reported in cells but its error is
governed by cell \emph{size}, which in three dimensions falls only as the cube
root of the count. The last step here raises the count by $1.55$ times, a
linear refinement of $1.157$; across the whole study, from $383\,616$ to
$1\,729\,232$ cells, the linear resolution improves by a factor of only
$1.65$. \textbf{A convergence study that looks like a fourfold refinement is a
sixty per cent one}, and a sidelobe that moves by a tenth of a decibel over it
has not been shown to be converging. In an explicit time-domain solver the time
step is set by the smallest cell, so the work goes as the four-thirds power of
the cell count and a genuine factor-of-two refinement would cost about sixteen
times the longest run in the study. The honest response is to quote a spread.

The same reasoning disqualifies the depth of the match from being quoted. It
improves monotonically across Table~\ref{tab:mesh} and therefore looks
converged, but what is converging is a null in a reflection coefficient---
another cancellation. What may be quoted is the fact of a match, not its depth.

\begin{table}[!t]
\renewcommand{\arraystretch}{1.15}
\caption{Mesh Convergence on the Production Geometry}
\label{tab:mesh}
\centering
\begin{tabular}{rcccc}
\hline
Cells & $f_{\mathrm{res}}$ & $|S_{11}|$ at & Peak & Sidelobe \\
 & / GHz & resonance / dB & / dBi & level / dB \\
\hline
$383\,616$ & $9.3583$ & $-26.12$ & $17.99$ & $-29.10$ \\
$776\,720$ & $9.3577$ & $-27.51$ & $18.07$ & $-29.69$ \\
$1\,117\,054^{\ast}$ & $9.3579$ & $-28.38$ & $18.08$ & $-29.95$ \\
$1\,729\,232$ & $9.3582$ & $-28.49$ & $18.09$ & $-29.84$ \\
\hline
\multicolumn{5}{l}{\footnotesize $^{\ast}$ The density everything else in this
paper was solved at.}
\end{tabular}
\end{table}

% =====================================================================
\section{Results: Where Pattern Multiplication Fails}
\label{sec:results}
\begin{figure*}[!t]
\centering
\includegraphics[width=0.94\textwidth]{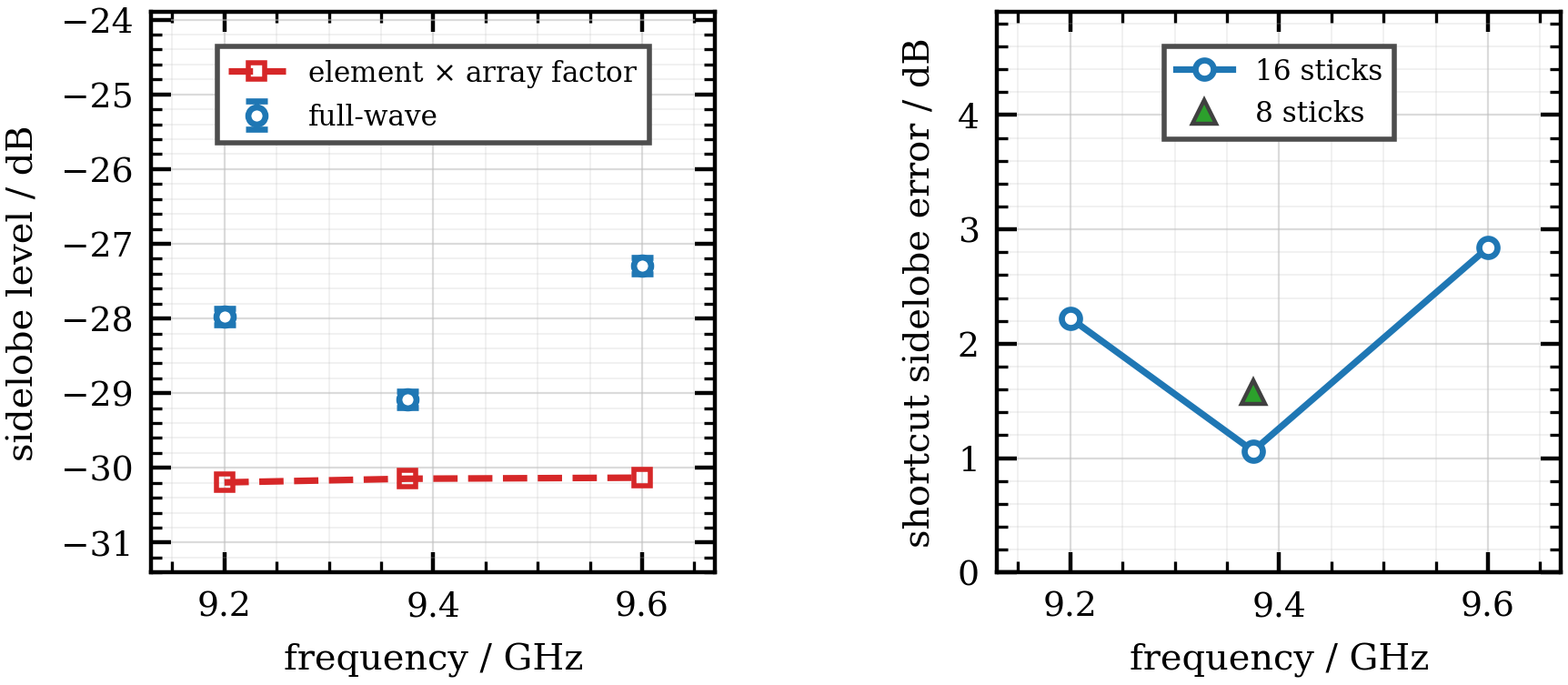}
\caption{The result. Left: sidelobe level of the sixteen-stick array in the
plane across the sticks---what element pattern $\times$ array factor predicts,
and what the full-wave solution of the same metal gives. The bars are the
$\pm0.11$~dB mesh spread of Section~\ref{sec:fw}, not a statistical confidence
interval. Right: the error, with the eight-stick array's error at the design
frequency for comparison. The penalty is not constant, and it is worst at the
band edges, where the design has least margin.}
\label{fig:shortcut}
\end{figure*}

% =====================================================================
The shortcut under test is to take the embedded pattern of one stick, multiply
it by the array factor of sixteen such sticks carrying the Taylor weights, and
call the product the array's pattern. Table~\ref{tab:shortcut} and
Fig.~\ref{fig:shortcut} put it beside the full-wave solution of the same metal
in the plane across the sticks.

\begin{table*}[!t]
\renewcommand{\arraystretch}{1.2}
\caption{Pattern Multiplication Against the Full-Wave Solution of the Same
Structure, in the Plane Across the Sticks}
\label{tab:shortcut}
\centering
\begin{tabular}{cccccc}
\hline
Frequency & Shortcut peak & Full-wave peak & Shortcut SLL & Full-wave SLL &
Sidelobe error \\
/ GHz & / dBi & / dBi & / dB & / dB & / dB \\
\hline
$9.200$ & $28.702$ & $30.736$ & $-30.194$ & $-27.970$ & $+2.224$ \\
$9.375$ & $29.442$ & $31.149$ & $-30.145$ & $-29.082$ & $+1.063$ \\
$9.600$ & $26.092$ & $29.765$ & $-30.131$ & $-27.290$ & $+2.841$ \\
\hline
\multicolumn{6}{l}{\footnotesize Every sidelobe figure in the last three
columns carries $\pm0.11$~dB of mesh spread (Section~\ref{sec:fw}).} \\
\multicolumn{6}{l}{\footnotesize Half-power beamwidths on the export grid are
$5.5^\circ$, $5.5^\circ$ and $5.0^\circ$.}
\end{tabular}
\end{table*}

Three statements follow, and each is a different kind of claim.

\emph{The error is of order one decibel and it is signed.} The shortcut is
optimistic everywhere---the real sidelobes are always higher than
predicted, by $+1.06$ to $+2.84$~dB. A designer who trusts it believes a
$-30$~dB array has been built when the article measures $-29.1$~dB at mid-band
and $-27.3$~dB at the top of it.

\emph{The error is not a constant that could be budgeted.} Its spread across
2\,\% of bandwidth is $1.78$~dB, larger than the error at the design frequency
itself. Adding a fixed decibel of margin to a shortcut calculation does not
make it safe; it makes it wrong in a different place. This is the practical
point of the paper: an error that were a stable bias would be a nuisance, and
one that is not is a hazard.

\emph{It does not vanish as the array grows.} The eight-stick array's error at
$f_0$ is $+1.58$~dB and the sixteen-stick array's is $+1.06$~dB. Doubling the
aperture recovers $0.51$~dB---the right sign and the right order for a fault
carried by the two edge sticks, whose number is fixed while the aperture
grows---but two points are not a curve, and no trend is drawn through them.
What the two points do establish is that the shortcut does not become safe
merely by building something bigger.

\subsection{The back lobe, where the shortcut has nothing to say}
In front-to-back ratio the shortcut and the solution are not close and are not
comparable: $-14.1$~dB against $-38.1$~dB. The shortcut's back lobe is a single
stick's back lobe, and a single stick is a narrow object that radiates freely
behind itself; sixteen of them side by side form a continuous conducting sheet.
The difference is $24.0$~dB. \textbf{This is not an error in the shortcut's
arithmetic; it is a structure the shortcut does not contain.}

The same thing happens in the plane containing the stick axes, where the
shortcut's claim is stronger still---it says the array's pattern in that plane
simply \emph{is} one stick's pattern, because the array factor is a constant
there. Measured, the array is $0.61$~dB worse in sidelobe level at $f_0$,
$6.82$~dB worse at $9.200$~GHz, and $27.5$~dB \emph{better} behind. Both
differences are the neighbours: coupling between sticks reweights the taper
inside each stick, and the sheet of metal the neighbours form closes the back.

\subsection{Why the coupling is large enough to do this}
The sixteen-port scattering matrix says why. A stick's nearest neighbour
receives $-15.2$~dB relative to what is fed to it, falling by about $6.4$~dB
for each further stick, against a self-reflection of $-27.0$~dB. \textbf{A
stick therefore delivers about $12$~dB more power into each of its neighbours
than it reflects to its own generator.} An array that is well matched port by
port can still be strongly coupled port to port, and the two facts are almost
independent; any argument of the form \emph{the array is well matched, so
coupling is small} is invalid, and this measurement is the counter-example.

Coupling \emph{within} a stick, by contrast, is negligible here: the evanescent
TE$_{20}$ field a slot leaves at its nearest neighbour, half a guide wavelength
away, is $-37.3$~dB below the field at the slot that launched it. That is the
quantity Elliott's internal-coupling correction is built to carry
\cite{elliott1978,elliott1979}, and for this geometry it does not need to be
applied. The coupling that matters is between sticks, through free space, and
it is exactly what the shortcut discards.

\subsection{The difference channel, reported as a bound}
\begin{figure*}[!t]
\centering
\includegraphics[width=0.94\textwidth]{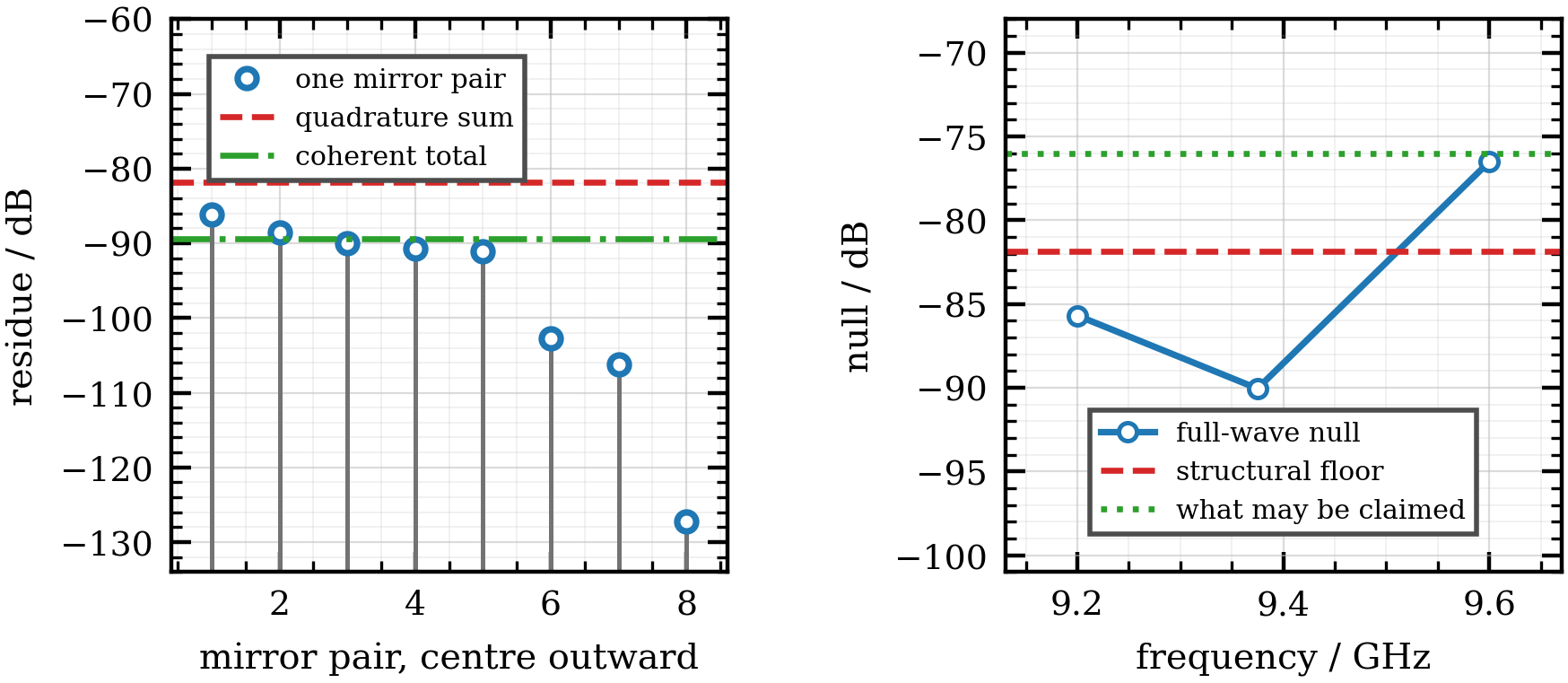}
\caption{The difference-channel null, reported as a bound. Left: the residue of
each of the eight mirror pairs at the design frequency, in decibels below the
\emph{sum}-channel peak, with their quadrature sum---the structural floor at
$-81.86$~dB---and their coherent total. The coherent total lies $7.54$~dB below
the quadrature sum, which means the pairs are cancelling each other and not
merely cancelling internally; that extra depth is an unprotected coincidence of
phase and is not claimed. Right: the null across the band against the
structural floor and against the figure this paper is willing to claim,
$-76$~dB. All values are referred to the sum-channel peak; referring them to
the difference channel's own peak makes them $3.99$~dB shallower.}
\label{fig:null}
\end{figure*}

Driving the two halves of the aperture in antiphase gives the monopulse
difference channel, whose boresight null is the most delicate quantity in the
design, because a near-perfect cancellation reports the error in its inputs at
full scale. Fig.~\ref{fig:null} shows the residue of each of the eight mirror
pairs at $f_0$ and the null across the band.

\textbf{What this paper claims is a bound: a structural floor of approximately
$-82$~dB below the sum-channel peak, and an achieved null below $-76$~dB across
$9.2$ to $9.6$~GHz.} The structural floor is the quadrature sum of the eight
pair residues, $-81.86$~dB, which is the depth the symmetry of the aperture
guarantees. The coherent total of the same eight residues is $7.54$~dB deeper
than the quadrature sum, but that extra depth is a coincidence of phase between
pairs and nothing in the design protects it, so it is not claimed. Nor is the
best value in band claimable: the null swings $13.54$~dB across $\pm2$\,\% in
frequency, and a quantity that moves that much with frequency is not a design
margin whatever its best value.

\textbf{Every null in this paper is referred to the sum-channel peak.} The
convention has to be stated because referring the same residue to the
difference channel's own peak makes it $3.99$~dB shallower; that figure is not
a convention either but the measured gap between the two peaks, $31.149$~dBi
against $27.150$~dBi. A $-76$~dB null referred to the sum peak demands that the
two halves of the aperture match to $0.0317$\,\% in amplitude and $0.0182^\circ$
in phase; the structural floor demands $0.0159$\,\% and $0.0091^\circ$. No
aperture that has been machined, brazed and plated holds those tolerances, so
\textbf{the full-wave null measures the symmetry of the model, not the symmetry
of the antenna}. What it does establish is that nothing in the design
itself---not the coupling, not the edge sticks, not the taper---sets a floor
above the bound quoted; the floor seen on a range will be set by the workshop,
and it is the workshop's number to report.

One method note, because it produced a wrong answer that looked like an antenna
result. A difference channel can be excited either by giving half the ports a
negative amplitude or by giving them a $180^\circ$ phase shift, and in a
frequency-domain post-processor these are not the same operation. A negative
amplitude is a true sign inversion at every frequency; a phase shift is
generally implemented as a time delay referred to a nominated reference
frequency, and at a reference of $9.5$~GHz, $180^\circ$ is $52.63$~ps, which is
$177.63^\circ$ at $9.375$~GHz---a residue of $2.4^\circ$ and a null floor near
$-34$~dB belonging to the dialogue box and not to the antenna. Every
difference-channel figure here was driven with signed amplitudes.

% =====================================================================
\section{Discussion}
\label{sec:discussion}
% =====================================================================
\subsection{What to use instead}
A result that demolishes a method owes the reader a replacement. There are
four, and the right one depends on how large the array is and on which quantity
is being claimed.

\emph{Element pattern $\times$ array factor} captures the aperture
distribution and the lattice, misses all coupling, the edges and the back, and
costs one stick and one excitation. That is the method measured here.

\emph{A periodic unit cell} captures the interior element's true environment
including coupling, misses the edge elements entirely and the finite aperture's
back face, and costs one cell and a scan sweep. \textbf{This is the option most
often mistaken for a full solution}: it is a genuine improvement on the
shortcut and it is still blind to exactly the elements this paper's error is
carried by. The evidence for reading the error as an edge effect is the pair of
measurements above---$+1.58$~dB on eight sticks and $+1.06$~dB on sixteen---
because a fault carried by a fixed number of edge elements dilutes as the
aperture grows and a fault carried by every element does not.

\emph{Interior and edge embedded patterns taken from a small array} capture
both environments, if the array is large enough that they are the only two, and
miss the slow variation between them. This is the working compromise: solve a
four- or eight-stick array, keep two embedded patterns rather than one, and
assemble the large array from those two. It removes the part of the error that
has been shown to dominate at a cost that does not grow with the size of the
array being designed. What it cannot do is tell you when it is wrong.

\emph{The whole aperture, every port driven} captures everything the geometry
contains, within the mesh, and costs an order of magnitude more cells and $N$
excitations. Because the third method cannot police itself, the full solve is
done once, at the end, on the article that is going to be cut.

\subsection{Scope of the claim}
This is one array, at one lattice, with one taper, in one band. The numbers
$+1.06$ to $+2.84$~dB belong to it and are not offered as a universal
correction. What generalises is weaker and more useful: that the error exists,
that it is signed the same way at every frequency measured, that it is
frequency-dependent by more than its own mid-band value, that it is carried by
the edge elements rather than distributed over the aperture, and that it is
larger than the numerical uncertainty of the solution that measures it by an
order of magnitude. A designer cannot take $1.06$~dB from this paper and add it
to their own shortcut calculation; what they can take is that the shortcut
needs auditing on their geometry, and the audit is the eight-versus-sixteen
comparison, which is cheap.

Two further limits should be stated. The lattice here is fixed at
$0.7943\lambda_0$ by the requirement that the sticks touch, which is dense; a
sparser lattice would couple less and might show a smaller error, and this was
not tested. And the calibration of Section~\ref{sec:cal} is valid over offsets
$0.61$ to $2.38$~mm in WR-90 at X band and nowhere else---a travelling-wave
taper in the same guide needs offsets beyond $3$~mm, and would need its
calibration extended before anything else.

\subsection{Why the calibration is a separate result}
The calibration is reported here as a supporting result, but it stands on its
own. The closed form for resonant length, (\ref{eq:Lclosed}), is printed in
handbooks with three terms of which only one depends on the design variable,
and that term moves the answer by $3.9~\mu$m over the range where the truth
moves $143.9~\mu$m. It is not that the expression is inaccurate; it is that it
contains no information about the quantity it is asked for, while looking as
though it does. Stevenson's conductance (\ref{eq:stevenson}) is the opposite
case and the more encouraging one: it is wrong by $5.16$\,\% and it is wrong by
$5.16$\,\% \emph{consistently}, to within $0.21$\,\%, over the whole useful
range, so a single measured constant repairs it while leaving intact the two
properties that make it worth keeping---the correct $g \to 0$ limit at zero
offset, and the frequency dependence carried entirely in $\lambda_g/\lambda_0$.
Correct a theory that carries the physics; do not replace it with a polynomial
that does not.

% =====================================================================
\section{Conclusion}
\label{sec:conclusion}
% =====================================================================
For a $16\times16$-slot planar slotted-waveguide array in WR-90 synthesised to
a $-30$~dB Taylor illumination and solved full-wave as a sixteen-port
structure, the element-pattern $\times$ array-factor shortcut underestimates
the sidelobe level by $+1.06$~dB at the design frequency and by $+2.22$ and
$+2.84$~dB at the edges of a 2\,\% band, each figure carrying $\pm0.11$~dB of
mesh spread. The error is signed, is worst where the design has least margin,
spreads $1.78$~dB across the band, and falls by only $0.51$~dB when the
aperture is doubled from eight sticks to sixteen---which places its origin in
the edge elements, where a unit-cell analysis is as blind as the shortcut. In
front-to-back ratio the two methods are not comparable at all, differing by
$24.0$~dB, because the shortcut does not contain the conducting sheet that
sixteen adjacent sticks form.

The array was made analysable by a thirteen-point single-slot full-wave
calibration using the reference-plane-independent extraction
$y = -2S_{11}/S_{21}$. That calibration replaced a handbook resonant-length
expression which moves $3.9~\mu$m over the design's offset range where the
truth moves $143.9~\mu$m, and it showed Stevenson's conductance to overestimate
by $5.16 \pm 0.21$\,\% over the useful range---wrong in value, right in shape,
and repairable by one constant. The difference channel of the same aperture is
reported as a bound: a structural floor of approximately $-82$~dB below the
sum-channel peak, achieved below $-76$~dB across $9.2$ to $9.6$~GHz, with the
reference peak stated because the two conventions differ by $3.99$~dB.

\section*{Acknowledgment}
AI assistance was used in preparing this manuscript.

\bibliographystyle{IEEEtran}
\bibliography{refs}

\end{document}